\documentclass[12pt]{article}
\usepackage{sigsam, amsmath,amsthm}
\usepackage{xcolor}
\usepackage{nicefrac}
\usepackage{enumerate}  
\newtheorem{theorem}{Theorem}
\newtheorem{proposition}[theorem]{Proposition}

\theoremstyle{definition}
\newtheorem{remark}[theorem]{Remark}
\usepackage{hyperref}

\issue{TBA}
\articlehead{TBA}
\titlehead{Factoring differential operators on algebraic curves in positive
characteristic}
\authorhead{Raphaël Pagès}
\begin{document}

\setlength{\abovedisplayskip}{.7ex}
\setlength{\belowdisplayskip}{.7ex}
\title{Factoring differential operators\\
over algebraic curves in positive characteristic}

\author{Rapha\"el Pag\`es \\
Institut de Math\'ematiques de Bordeaux\\
University of Bordeaux \\
Talence, France, 33 405 \\
\url{https://www.math.u-bordeaux.fr/~rpages/}}

\date{May 20, 2022}

\maketitle

\begin{abstract}
We present an algorithm for factoring linear differential operators with
coefficients in a finite separable extension of $\mathbb{F}_p(x)$.
Our methods rely on specific tools arising in positive characteristic:
$p$-curvature, structure of simple central algebras and $p$-Riccati
equations.
\end{abstract}

\section{Introduction}

Studying and solving differential equations has been an important subject
on mathematicians' mind since the invention of differential calculus and
has found many applications. Although those equations are generally studied
on real or complex variables, there is an algebraic counterpart to this
theory, which makes sense over any base field, including number fields,
$p$-adic fields and fields of positive characteristic. Applications 
include points counting on elliptic curves~\cite{Lauder04}, isogeny 
computations~\cite{LaVa16,Eid21} and, more generally, the study of (the 
cohomology of) many arithmetic varieties.

In this work, we focus on linear differential equations of the form $L(y)=0$
where
\[L= a_r(x)\partial^r+ a_{r-1}(x)\partial^{r-1}+\cdots+a_1(x)\partial+a_0(x)\]
and the $a_i(x)$ are regular functions on an algebraic curve. The variable 
$\partial$ acts by derivation and $L$ is thus a differential operator. 
The set of differential operators is provided with a ring structure 
derived from the Leibniz rule. A natural question arising when studying 
linear differential operators is that of factorisation.

The case of 
operators with coefficients in $\mathbb{C}(x)$ is well understood and 
several algorithms have been proposed throughout the years~\cite{Gri90,
VHo97,ChGoMe22}. They usually rely on transcendental arguments, \emph{e.g.} on
properties of the monodromy group.
In characteristic $p$, the monodromy does not exist but other powerful
tools are available. One of them is the $p$-curvature: it was used in
the context of factorisation for the first time by van der 
Put~\cite{Put95,Put96}. In his PhD thesis, Cluzeau developed this
approach and described a factorisation algorithm for linear
differential systems over $\mathbb{F}_q(x)$ (where~$\mathbb{F}_q$ is a
finite field of characteristic $p$)~\cite{Cluzeau03,Cluzeau04}.
In this work, we present an algorithm that completely factors any 
differential operator with coefficients in a finite separable extension 
$K$ of $\mathbb{F}_p(x)$.

\section{Main ingredients}
Let $K$ be a finite separable extension of $\mathbb{F}_p(x)$. The natural
derivation $\frac{\mathrm{d}}{\mathrm{d} x}$ extends uniquely to $K$ and
we let $K\langle\partial\rangle$ denote the ring
of linear differential operators with coefficients in $K$.
For $L\in K\langle\partial\rangle$, we set
$\mathcal{D}_L:=\nicefrac{K\langle\partial\rangle}{K\langle\partial\rangle
L}$. Here are the main ingredients that we will be using in our algorithm:
\begin{itemize}\setlength\itemsep{.2ex}
  \item[(I1)]the one-to-one decreasing bijection between
the set of right divisors of $L$ (up to a multiplicative element of
$K^\times$) and the set of $K\langle\partial\rangle$-submodules of
$\mathcal{D}_L$ given by $$L'\mapsto \mathcal{D}_L
L':=\nicefrac{K\langle\partial\rangle L'}{K\langle\partial\rangle L};$$
this bijection also induces nice relations between the sum and intersection
of submodules, and the greatest common right divisor and least common left
multiple of operators respectively,
  \item[(I2)]the $p$-curvature of $L$ which will allow us to
find a first factorisation of $L$ as a product of operators verifying
additional properties,
  \item[(I3)]the arising central simple algebra structure and the Morita equivalence
which will allow us to rephrase our problem through the prism of linear algebra
and eventually reduce it to solving a ``$p$-Riccati''
equation,
  \item[(I4)]tools of algebraic geometry such as the Jacobian of an
    algebraic curve to solve this equation.
\end{itemize}

\section{Using the $p$-curvature}

For any $f\in K$, $\frac{\mathrm{d} }{\mathrm{d} x}f^p=0$. The set of
elements of the form $f^p$ forms the subfield of constants of $K$ which
we denote by $C$. Additionally for any $f\in K$, $\left(\frac{\mathrm{d}}{\mathrm{d} x}\right)^p f=0$. Thus the left
multiplication by $\partial^p$ induces a $K$-linear endomorphism of
$\mathcal{D}_L$: it is the so-called ``$p$-curvature'', which we
denote by $\psi^L_p$.  Its characteristic polynomial
$\chi(\psi^L_p)$ has coefficients in $C$.
We factor $\chi(\psi^L_p)=\prod_{i=1}^n N_i^{\nu_i}(Y)$ in the ring
$C[Y]$ (commutative factorisation) where the $N_i(Y)$ are pairwise
distinct irreducible polynomials over $C$.
The kernel decomposition lemma states:
\[\mathcal{D}_L=\bigoplus_{i=1}^n \ker N_i^{\nu_i}(\psi^L_p).\]
Applying (I1), this decomposition translates to a
first factorisation of $L$:
\begin{theorem}
  There exists a factorisation $L = L_1 \cdots L_n$ such that
  $\chi(\psi^{L_i}_p)=N_i^{\nu_i}(Y)$ for all $i \in \{1, \ldots, n\}$
  and $L_n=\mathrm{gcrd}(L,N_n^{\nu_n}(\partial^p))$.
\end{theorem}

\begin{remark}
  Since $\mathcal{D}_L$ decomposes as a direct sum
  of submodules, we even get a $\mathrm{lclm}$ factorisation:
  $L=\mathrm{lclm}_{i=1}^n(\mathrm{gcrd}(L,N_i^{\nu_i}(\partial^p))).$
\end{remark}

From what precedes, we can safely suppose that $\chi(\psi^L_p)$ is a 
power of an irreducible polynomial in $C[Y]$ of the form $N^\nu(Y)$.
By recursively considering the $\mathrm{gcrd}(L,N(\partial^p)^i)$,
we can even further assume that $L$ is a
divisor of $N(\partial^p)$ for some irreducible polynomial $N$ in
$C[Y]$.

\section{Factorisation of central irreducible elements}

Let $L\in K\langle\partial\rangle$ be a divisor of some $N(\partial^p)$
with $N$ irreducible in $C[Y]$. The quotient $\mathcal{D}_L$ has a structure of a
$\mathcal{D}_{N(\partial^p)}$-module. Write $C_N=\nicefrac{C[Y]}{(N)}$; it
is a field extension of $C$. Let $y_N$ be the image of $Y$ in $C_N$. To avoid technicalities, we
shall assume that $C_N$ is separable. 
We set $K_N=K\cdot C_N$.
\begin{theorem}[\cite{Revoy73,Put95,BoCaSc14}]
  The quotient ring $\mathcal{D}_{N(\partial^p)}$ is a simple central
  algebra over $C_N$.
\end{theorem}

Using the Artin-Wedderburn theorem~\cite[Thm.~2.1.3]{AnFu92},
one shows that $\mathcal{D}_{N(\partial^p)}$ is either a division
algebra or isomorphic to $M_p(C_N)$ (the ring of $p \times p$ matrices
over $C_N$).
In the
former case, $\mathcal{D}_{N(\partial^p)}$ has no nontrivial zero divisor,
meaning that $N(\partial^p)$ itself is irreducible.

Let us now suppose that $\mathcal{D}_{N(\partial^p)}$ is a matrix algebra.
The Morita equivalence~\cite[\S6]{AnFu92} provides us with a (nonexplicit) 
decreasing bijection 
between submodules of $\mathcal{D}_{N(\partial^p)}$ and
sub-$C_N$-vector spaces of $C_N^p$.
Furthermore, if $N(\partial^p)$ factors as $LL'$ then $\mathcal{D}_L$ is
identified with
$\mathcal{D}_{N(\partial^p)}L'\subset\mathcal{D}_{N(\partial^p)}$. We write~$V$ for the corresponding subspace of $C^p_N$. Combining Morita equivalence
with (I1), we conclude that
irreducible divisors of $L$ are in one-to-one correspondence with
hyperplanes of $V$. Those can be found
by computing the intersections of $V$ with generic hyperplanes of
$C^p_N$. Specifically, what we need is a family of $p$ hyperplanes of 
$C^p_N$ whose intersection is reduced to zero, which in turn corresponds
to finding a factorisation of $N(\partial^p)$ as an $\mathrm{lclm}$ of 
irreducible differential operators.

There is now an isomorphism
$\varphi_N:\large{\nicefrac{K\langle\partial\rangle}{(N(\partial^p))}\overset{\sim}{\longrightarrow}
\nicefrac{K_N\langle\partial\rangle}{(\partial^p-y_N)}}$.
Thus finding irreducible divisors of $N$ amounts to finding
irreducible divisors of $\partial^p-y_N$ with coefficients in $K_N$.
Such divisors are of the form $\partial-f$, with $f\in K_N$ verifying the
following ``$p$-Riccati'' equation:
\begin{equation}\label{p-Riccati}f^{(p-1)}+f^p=y_N.\end{equation}
  We let $\mathcal{S}_N$ be the set of solutions of~\eqref{p-Riccati}. 
  It turns out that $\mathcal{S}_N$ can be fully obtained from a
  particular solution by adding logarithmic derivatives of
  functions in $K_N$.

\begin{theorem}
  Set
  $\mathcal{L}_f:=\mathrm{lclm}\big(\mathrm{gcrd}\big(N(\partial^p),\,\varphi_N^{-1}(\partial{-}f)\big),\,L'\big)\cdot
  L^{'-1}$.
  \begin{enumerate}[i)]\setlength\itemsep{.2ex}
    \item If $L=N(\partial^p)$ then $f \mapsto \mathcal{L}_f$ is a
    one-to-one correspondence between $\mathcal{S}_N$ and the set of irreducible
       right divisors of $N(\partial^p)$.
     \item In general, all irreducible right divisors of $L$ are of the
       form $\mathcal{L}_f$ with $f\in\mathcal{S}_N$.
     \item For all $f\in \mathcal{S}_N$,
       $L=\mathrm{lclm}\big(\mathcal{L}_{f},
        \mathcal{L}_{f+\frac{1}{x}}, \ldots,
        \mathcal{L}_{f+\frac{p-1}{x}}\big)$.
  \end{enumerate}
\end{theorem}

\section{Resolution of the ``$p$-Riccati'' equation}

In~\cite[\S 13.2.1]{PuSi03}, Singer and van der Put explain how to solve the
$p$-Riccati equation over $\mathbb{F}_q(x)$. The idea is somehow to 
show that if Eq.\eqref{p-Riccati} has a solution, then it has another
solution with at most the same poles as $y_N$. 
We then deduce a bound on the degree of the numerator and conclude using
$\mathbb{F}_p$-linearity.
The method for the general case follows the same pattern. However, in
full generality, all solutions may have more poles than $y_N$.
In order to get around this issue, we use tools from algebraic geometry:
Riemann-Roch spaces, Picard group of a curve and Jacobians.
For $f \in K_N$, let $\nu_{\mathfrak P}(f)$ denote the order of vanishing
of $f$ at the place~$\mathfrak P$.

\begin{proposition}
Let $\mathfrak{P}$ be a place of
$K_N$ and $t_\mathfrak{P} \in K_N$ such that $\nu_{\mathfrak{P}}(t_\mathfrak{P})=1$.
Let $f$ be a solution of Eq.~\eqref{p-Riccati}. Then
  $\nu_{\mathfrak{P}}(f)\geq\min(0,p^{-1}\nu_{\mathfrak{P}}(y_N),\nu_{\mathfrak{P}}(t_\mathfrak{P}')-1)$.
Besides, when $\mathfrak{P}$ is not a pole of $y_N$ 
nor a ramified place, nor a place at infinity, the residue of $f$ at 
$\mathfrak{P}$ (denoted by $\mathrm{Re}_\mathfrak{P}(f)$)
is an integer.
\end{proposition}

It follows from the previous proposition that, when $\mathfrak{P}$ is not a 
pole of $y_N$, nor a ramified place, nor a place at infinity, we always 
have $\nu_{\mathfrak{P}}(f)\geq -1$. Moreover, when equality holds,
one can remove the simple pole at $\mathfrak{P}$ by replacing $f$ by
$f- g'/g$ where $g \in K_N$ has a zero of 
order $\mathrm{Re}_\mathfrak{P}(f)$ at $\mathfrak{P}$.
Unfortunately, this transformation may lead to other undesirable poles.
In order to control this back-and-forth, we use computations in the group 
of divisors 
of $K_N$ (which is, by definition, the free commutative group generated 
by the set of places of $K_N$).

\begin{proposition}
\label{prop:places}
  Let $S$ be a set of places containing the poles of $y_N$, the ramified
  places of $K_N$ and the places at infinity. Let $\mathfrak S$ be a
  fixed place of $K_N$ of degree~$1$. Set
  $D=c\cdot \mathfrak{S}+\sum_{\mathfrak{P}\notin S} \mathrm{Re}_\mathfrak{P}(f)\cdot\mathfrak{P}$
  where $c\in\mathbb{Z}$ is such that $\deg(D)=0$. If
  there exist two divisors $D_p$ and $D'$ such that
  $D - D' - p D_p$ is a principal divisor, then Eq.~\eqref{p-Riccati} 
  has a solution with no pole outside $S \cup D' \cup \{\mathfrak S\}$. 
\end{proposition}

The group of divisors of $K_N$ of degree $0$ modulo the subgroup of
principal divisors is the so-called \emph{Picard group} of $K_N$ and
is denoted by $\text{Pic}^0(K_N)$.
Proposition~\ref{prop:places} above ensures that we will get an explicit 
bound on the poles of a solution of Eq.~\eqref{p-Riccati} if we can
bound the cokernel of the multiplication by $p$ on 
$\text{Pic}^0(K_N)$. This finally can be achieved using general
results on the Jacobian of $K_N$.


{\small
\begin{thebibliography}{10}
\renewcommand{\itemsep}{0pt}

\bibitem{AnFu92}
F.~W. Anderson and K.~R. Fuller.
\newblock {\em Rings and categories of modules}, volume~13 of {\em Graduate
  Texts in Mathematics}.
\newblock Springer-Verlag, New York, second edition, 1992.

\bibitem{BoCaSc14}
A.~Bostan, X.~Caruso, and E.~Schost.
\newblock A fast algorithm for computing the characteristic polynomial of the
  p-curvature.
\newblock In {\em I{SSAC} 2014---{P}roc. of the 39th {I}nternational
  {S}ymposium on {S}ymbolic and {A}lgebraic {C}omputation}, pages 59--66. ACM, 2014.
       
\bibitem{ChGoMe22}
F.~Chyzak, A.~Goyer, and M.~Mezzarobba.
\newblock {Symbolic-Numeric Factorization of Differential Operators}.
\newblock To appear in {\em I{SSAC} 2022}.

\bibitem{Cluzeau03}
T.~Cluzeau.
\newblock Factorization of differential systems in characteristic {$p$}.
\newblock In {\em Proc. of the 2003 {I}nternational {S}ymposium on
  {S}ymbolic and {A}lgebraic {C}omputation}, pages 58--65. ACM, 2003.

\bibitem{Cluzeau04}
T.~Cluzeau.
\newblock {\em Algorithmique modulaire des \'equations diff\'erentielles
  lin\'eaires}.
\newblock PhD thesis, Universit\'e de {L}imoges, 2004.

\bibitem{Eid21}
E.~Eid.
\newblock Fast computation of hyperelliptic curve isogenies in odd
  characteristic.
\newblock In {\em I{SSAC} 2021---{P}roc. of the 39th {I}nternational
  {S}ymposium on {S}ymbolic and {A}lgebraic {C}omputation}. ACM,
  2021.

\bibitem{Gri90}
D.~Y. Grigoriev.
\newblock Complexity of factoring and calculating the {GCD} of linear ordinary
  differential operators.
\newblock {\em J. Symbolic Comput.}, 10(1):7--37, 1990.

\bibitem{LaVa16}
P.~Lairez and T.~Vaccon.
\newblock On $p$-adic differential equations with separation of variables.
\newblock In {\em I{SSAC} 2016---{P}roc. of the 39th {I}nternational
  {S}ymposium on {S}ymbolic and {A}lgebraic {C}omputation}, pages 319--323.
  ACM, 2016.

\bibitem{Lauder04}
A.~G.~B. Lauder.
\newblock Deformation theory and the computation of zeta functions.
\newblock {\em Proc. London Math. Soc. (3)}, 88(3):565--602, 2004.

\bibitem{Revoy73}
P.~Revoy.
\newblock Alg\`ebres de {W}eyl en caract\'{e}ristique {$p$}.
\newblock {\em C. R. Acad. Sci. Paris S\'{e}r. A-B}, 276:A225--A228, 1973.

\bibitem{Put95}
M.~van~der Put.
\newblock Differential equations in characteristic {$p$}.
\newblock volume~97, pages 227--251. 1995.
\newblock Special issue in honour of Frans Oort.

\bibitem{Put96}
M.~van~der Put.
\newblock Reduction modulo {$p$} of differential equations.
\newblock {\em Indag. Math. (N.S.)}, 7(3):367--387, 1996.

\bibitem{PuSi03}
M.~van~der Put and M.~F. Singer.
\newblock {\em Galois theory of linear differential equations}, volume 328 of
  {\em Grundlehren der Mathematischen Wissenschaften [Fundamental Principles of
  Mathematical Sciences]}.
\newblock Springer-Verlag, Berlin, 2003.

\bibitem{VHo97}
M.~van Hoeij.
\newblock Factorization of differential operators with rational functions
  coefficients.
\newblock {\em J. Symbolic Comput.}, 24(5):537--561, 1997.

\end{thebibliography}}

\end{document}